\documentclass{phbauth}       
\usepackage{graphicx}

\begin{document}

\begin{frontmatter}

\title{Flux lattice melting and the onset of $H_{c2}$ fluctuations}

\author[address1]{Stephen W.
Pierson\thanksref{thank1}}
\author[address2]{Oriol T. Valls}

\address[address1]{Department of Physics, Worcester Polytechnic Institute,
Worcester, MA 01609-2280, USA} \address[address2]{ 
Physics Department 
and Minnesota Supercomputer Institute,
University of Minnesota, Minneapolis, MN 55455, USA}

\thanks[thank1]{Corresponding
author. E-mail: pierson@wpi.edu; FAX: (508) 831-5391}

\begin{abstract}
The flux lattice melting temperature in YBa$_2$Cu$_3$O$_{7-\delta}$ has
been shown to be very close to that of the onset of fluctuations around
$H_{c2}(T)$. Here, we present a theoretical argument in support of the
idea that this occurs because the increased strength of the fluctuations
as a function of magnetic field pushes away the first order flux lattice
melting transition. The argument is based on hydrodynamic considerations
(the Hansen-Verlet freezing criterion). It is not specific to
high-temperature superconductors and can be generalized to other systems.
\end{abstract}

\begin{keyword}
Flux Lattice Melting; Hansen-Verlet freezing criterion; fluctuations
\end{keyword}

\end{frontmatter}

A prominent feature of the remarkably
rich field-temperature ($H$-$T$) phase diagram in 
high-T$_c$ 
superconductors\cite{blatter94} (HTSC's) is the flux 
lattice melting (FLM) temperature $T_M(H)$ line. 
Theories of $T_M(H)$ typically
treat that phase transition as a stand-alone phenomenon, as in, 
for example, the earliest approach of examining the elastic moduli of the
flux lattice to establish the Lindemann criterion for the melting
temperature.\cite{houghton89} 

In contrast, we have 
recently presented evidence\cite{pierson98} that 
$T_M(H)$ should be viewed as intimately connected 
to a nearby feature in the phase diagram: the 
superconducting/normal crossover line 
$H_{c2}(T)$. In Ref.~\cite{pierson98}, it was shown 
that the FLM temperature coincides with that of the 
onset of fluctuations around $H_{c2}$, as the
temperature is increased. The essential idea, which is
appealing in its simplicity, is best captured
by considering the system as it is being cooled: as 
one does so, the vortices cannot freeze into a lattice 
until the $H_{c2}$ fluctuations have died down. When 
the strength of the fluctuations is estimated through 
the field-dependent Ginzburg criterion $Gi(H)$, the
freezing temperature of the vortices is determined by, 
\begin{equation} 
Gi(H)=[T_m(H)-T_c(H)]/T_c(H)\propto H^{2/3}, 
\label{ginzH}
\end{equation} 
where the constant of proportionality is presented in
Refs.~\cite{pierson98} and \cite{herbut95}.

In this paper, we present a simple theoretical 
argument, based on very general hydrodynamic
considerations and the Hansen-Verlet\cite{hv} (HV)
freezing criterion (the freezing version
of the better-known Lindemann melting rule),
that explains the coincidence
of the FLM with the edge of the $H_{c2}$
fluctuation region. 

The evidence for the relation between the FLM 
and $H_{c2}$ fluctuation lines is
twofold.\cite{pierson98} The fit of the theoretical
expressions for the specific heat of Te\u sanovi\' c and 
Andreev\cite{zlatko94} to the YBa$_2$Cu$_3$O$_{7-\delta}$ 
(YBCO) data of Schilling {\it et al.}\cite{schilling97} 
(see Fig.~1), establishes that the specific heat for temperatures 
above the FLM (where it has a spike) can be 
attributed to $H_{c2}$ 
fluctuations and that the spike coincides with the onset of those
fluctuations. Secondly, the positions of the FLM specific
heat spikes are determined by fluctuations around $H_{c2}$ as described by
the formula of Herbut and Te\u sanovi\' c.\cite{herbut95} [The formula is
essentially Eq.~(\ref{ginzH}).]

\begin{figure}[b]
\begin{center}\leavevmode
\includegraphics[width=0.93\linewidth]{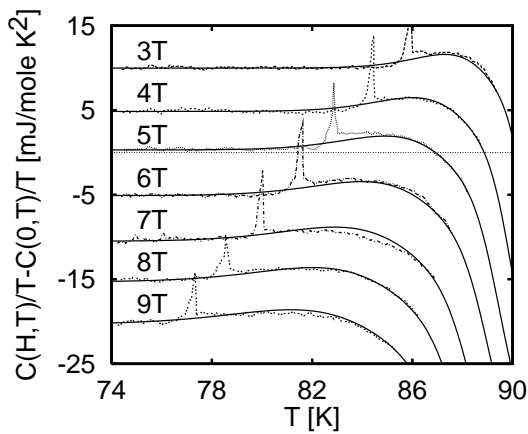}
\caption{ 
The fit of the Te\u sanovi\' c-Andreev\cite{zlatko94} 
GL-LLL expressions to the Schilling {\it et al.} 
specific heat YBCO data\cite{schilling97}.
}\label{compare1}\end{center}\end{figure}

Here we argue on very simple and general theoretical 
grounds that a first order freezing transition 
cannot occur until the fluctuations
from a nearby second order transition have  
subsided. We use only two fundamental hydrodynamic
relations and the Hansen-Verlet 
criterion. The HV criterion states 
that freezing occurs when, as one
cools, the magnitude of the first finite wavevevector
peak in the static structure factor
$S(k)$ reaches a certain value, typically 
ranging from 3 to 6. There are two hydrodynamic 
constraints\cite{by}: the long wavelength limit of
$S(k)$ is related to the average
density $\rho$ and the
compressability $\chi$ by 
\begin{equation}
\lim_{k\rightarrow 0}S(k)=k_BT\rho\chi, 
\label{limit} 
\end{equation} 
while on the other hand,
the integral of $S(k)-1$ over all $k$ varies 
with $T$ only  
very slowly, through $\rho$. 

Combining the two hydrodyamic results with the HV 
criterion, one immediately sees that the fluctuations
arising from the higher temperature second order
transition must have died down before the first order 
transition may occur. Just below  the second order phase 
transition, where $\chi$ diverges, $S$ still has
a large peak at small $k$. Because of the constraint
on the integral over all $k$ of $S(k)$ the magnitude 
of the finite $k$ peaks must remain  fairly 
small. Only as one moves further below the second
order phase transition, and the zero $k$ peak shrinks, 
can the magnitude of the first peak increase to a value 
large enough to 
satisfy the Hansen-Verlet rule. Hence, freezing can occur 
only sufficiently far from the second order phase transition.
Simple free energy models involving two coupled order
parameters can be built to illustrate this conclusion.

Given the generality of this argument, it is clear that it 
applies to any system where a first order
phase transition is near a second order one. Can\-di\-dates
include liquid crystals, heavy-fermion superconductors, 
materials with structural and ferroelectric
transitions, and Langmuir monolayers. 

We have shown that FLM in
the $H$-$T$ phase diagram of HTSC's cannot
occur until the fluctuations from the $H_{c2}(T)$  line die
down, evidence for which was given in Ref.~\cite{pierson98}.
Our theory is sufficiently general that it 
can be applied to other systems.

We thank I. Herbut and Z. Te\u sanovi\' c for discussions.
SWP thanks the  
Petroleum Research Fund, administered by the ACS, for 
their support.

\end{document}